# Scaling properties of soft matter in equilibrium and under stationary flow

Armando Gama Goicochea

División de Ingeniería Química y Bioquímica, Tecnológico de Estudios Superiores de Ecatepec, Av. Tecnológico s/n, Ecatepec, Estado de México 55210, Mexico
`agama@alumni.stanford.edu`

**Abstract.** A brief review is presented of the scaling of complex fluids, polymers and polyelectrolytes in solution and in confined geometry, in thermodynamical, structural and rheology properties using equilibrium and non – equilibrium dissipative particle dynamics simulations. All simulations were carried out on high performance computational facilities using parallelized algorithms, solved on computers using both central and graphical processing units. The scaling approach is shown to be a unifying axis around which general trends and basic knowledge can be gained, illustrated through a series of case studies.

**Keywords:** Scaling, polymers, polyelectrolytes, radius of gyration, Couette flow, viscosity, friction coefficient

## 1 Introduction

Scaling is one of the most cherished concepts in physics and its application to soft matter has been as successful as it has in other areas of physics. At the heart of it is the idea that if a system is self – similar and it is not under the influence of long range interactions, then it should display properties whose general behavior is invariant under scale transformations. A superb account of it is de Gennes's treatise [1], where he uses simple concepts to arrive at profound and general scaling laws for polymers under various circumstances. This could be accomplished in part because of the advanced state of the experimental efforts to understand the nature of complex fluids. Searching for scaling is important also from a practical perspective, because quantitative predictions can be made about systems of vastly different chemical composition.

At the beginning of the seventies of the past century there were also fundamental theoretical developments, such as the renormalization group [2] and fractal geometry [3] later, which led to the robust understanding of scaling particularly, in physics. At the time reference [1] was written, molecular simulation was still a novel tool and there were only a few works focused on testing scaling laws. The advent of modern computers, with fast processors, efficient architecture and optimum algorithms has made of molecular simulation an indispensable research tool, one that has become commonplace both in academia and in the productive sector as well [4].

When testing scaling laws using molecular simulation one faces the challenge of reducing finite size effects that could potentially mask the underlying scaling phenomena. Doing so means performing numerical simulations on systems of increasing size, so that the phenomenon under study can be traced over considerable changes of scale, which in turn requires longer simulation time. If one uses atomistically detailed models [5], this task quickly becomes prohibitive or impractical due to the relatively long range of basic interaction models such as the Lennard – Jones potential. Since all simulations are of finite size, one must cut the range of the interactions using some criterion, which may lead to artifacts unless large systems are used [5].

In addition to the so – called long range corrections employed when the interactions are long – ranged [5], there is also a systematic approach that helps simulate large systems using moderately sized simulation boxes. Such approach, generally termed "coarse – grained" models [6], typically consists of integrating out some degrees of freedom, yielding effective interactions that can be thought of as potentials of mean force rather than basic interactions. What one loses in atomic – scale detail is gained in mesoscopic – scale information and in savings in computational time. Among the most successful and popular coarse – grained methods is the one known as dissipative particle dynamics [7] (DPD), introduced originally to study the rheology of colloidal suspensions. Once its statistical mechanics foundations were correctly laid out [8], the potential and versatility of DPD was quickly recognized and it became widely applied to model systems as diverse a proteins [9], surfactants and polymers in solution [10], biological membranes [11], paints [12] and even to phenomena such as thrombosis [13]. The DPD model is now well known among practitioners of numerical simulation; there are also various reviews available that detail its foundations and some of its most successful applications [14 – 16]. Therefore, for the sake of brevity, only what is pertinent shall be presented here; the reader is referred to the cited reviews for additional details.

In this Chapter I revisit some of the recent work carried out by our group on scaling properties of soft matter systems such as polymers in solution and under confinement, polymer brushes and polyelectrolytes in equilibrium and under stationary flow, using DPD. In Section 2 the essentials of the DPD model are presented briefly, followed in Section 3 by the scaling of the interfacial tension in mixtures of organic liquids and water. Section 4 is devoted to the scaling of polyelectrolytes under different solvent conditions, while Section 5 focuses on various scaling features of polymer brushes. Results on the scaling of polymer brushes under flow can be found in Section 6. In Section 7 the scaling of polymers in two dimensions are revisited. The conclusions are laid out in Section 7. Emphasis is placed on the discussion of physical ideas; for simulation details and additional information the reader is referred to the original articles.

## 2 Models and Methods

The DPD model is based on the integration of the internal degrees of freedom of groups of atoms used to construct the DPD particles or beads whose motion is solved through the integration of Newton's second law of motion, following an algorithm that is essentially the same as the one used in atomistic simulations [5]. The conservative interaction between DPD particles is a linearly decaying, short range force:

$$\mathbf{F}_{ij}^C = \begin{cases} a_{ij}(1 - r_{ij})\hat{\mathbf{r}}_{ij} & r_{ij} \leq r_c \\ 0 & r_{ij} > r_c \end{cases} \quad , \tag{1}$$

where $a_{ij}$ is a strength of the interaction, which determines the thermodynamics of the system, and $r_C$ is the cutoff distance that sets the length scale of the interactions. Notice that the force remains finite even when the centers of mass of two interacting particles overlap, meaning that DPD particles can in principle occupy the same space at the same time. However, this does not occur in practice because the interaction strength $a_{ij}$ is at least $25k_BT/r_C$ or larger, therefore there is a very large energy cost involved in the full overlap of DPD particles. This was not originally recognized when electrostatic interactions were introduced into the DPD model [17], as it was believed that the soft nature of the DPD beads containing charges would lead to the formation of ionic pairs of infinite electrostatic energy, which is of course unphysical. In Section 4 I show this is not really a problem, and point charges can in fact be used in DPD leading to correct and artifact – free conclusions.

The short range nature of the force in eq. (1) is the key to the mesoscopic reach of DPD and its capability of produce simulations with observation times of the order of tens of microseconds, setting it at least three orders of magnitude apart from its all – atom counterparts [4]. It is also the reason why one can do away with long range correction and why finite size effects in DPD are minimal [18]. Yet, what is perhaps more advantageous is the DPD thermostat: the local viscosity of the fluid is modeled as a dissipative force, whose energy dissipation is invested into local Brownian motion, modeled by a random force. As is customary when these types of forces are present, one must make sure that the fluctuation – dissipation theorem is obeyed; doing so in DPD leads to a relation between the strengths of the dissipative ($\gamma$) and random ($\sigma$) forces given by $k_BT = \frac{\sigma^2}{2\gamma}$, which sets up the thermostat [8]. All forces are pairwise additive, leading to global momentum conservation. The conservative ($\mathbf{F}^C$), dissipative ($\mathbf{F}^D$), and random ($\mathbf{F}^R$) forces acting between any two particles $i$ and $j$, placed a distance $r_{ij}$ apart must be integrated in finite time steps to yield the momenta and positions of all particles:

$$\dot{\mathbf{p}}_i = \sum_{j \neq i} \mathbf{F}_{ij}^C + \sum_{j \neq i} \mathbf{F}_{ij}^D + \sum_{j \neq i} \mathbf{F}_{ij}^R \quad . \tag{2}$$

All forces between particles $i$ and $j$ are zero beyond a finite cutoff radius $r_c$, which is usually also chosen as $r_c \equiv 1$. The natural probability distribution function of the DPD model is found to be that of the canonical ensemble [8], where $N$ (the total particle number), $V$, and $T$ are kept constant, although it is equally possible to solve the system using the Monte Carlo algorithm [4] under various ensembles of interest [18 – 20]. Polymer chains can be constructed following the Murat – Grest bead – spring linear chain model [21], while surfaces can either be introduced using effective force fields [19, 22] or by freezing layers of particles [23]. The chemical composition of the DPD particles is incorporated into the value chosen for the conservative interaction strength, $a_{ij}$, see eq. (1), usually obtained from the Flory – Huggins solution theory [24]. Full details and several applications of DPD can be consulted in recently published reviews [14 – 16].

## 3   Interfacial tension scaling

The pioneering work of Widom and coworkers [25, 26] established that the interfacial tension between two liquids at finite temperature, $\sigma(T)$, could be expressed as

$$\sigma(T) = \sigma_0 \left(1 - \frac{T}{T_C}\right)^\mu, \tag{3}$$

where $T_C$ is the critical temperature, at which the interface becomes unstable, $\sigma_0$ is a system – dependent constant, and $\mu$ is the scaling exponent, whose currently accepted value is $\mu = 1.26$ [27]. On the other hand, the correlation length of the phases, $\xi$, ignoring logarithm corrections, can be written as

$$\xi(T) = \xi_0 \left(1 - \frac{T}{T_C}\right)^{-\nu}, \tag{4}$$

where $\xi_0$ is also system depending and $\nu$ is the scaling exponent. Its value for the three – dimensional Ising model is $\nu = 0.63$ [28]. The energy of the liquid mixture, $k_B T$, can be expressed as the product of the interfacial tension times the area defined by the correlation length, which is generalized in $d$ dimensions as $\xi^{d-1}$; mathematically, $k_B T \sim \sigma(T)\xi^{d-1}$. As the system approaches its critical point $k_B T \to k_B T_C$, which must be temperature – independent; combining then eqs. (3) and (4) yields the following hyperscaling relationship between the scaling exponents in those equations [26]:

$$\mu = (d-1)\nu. \tag{5}$$

To test eq. (5) one must first device a model to introduce the temperature dependence into the DPD framework. The first work to accomplish that is due to Mayoral and Gama Goicochea [29], where temperature changes are introduced through the temperature dependence of the conservative interaction strength, $a_{ij}$, see eq. (1). The dependence of $a_{ij}$ on temperature is in turn obtained from the dependence of the Flory – Huggins parameter on temperature by means of the solubility parameters. Equation (5) was tested in 3$d$ for mixtures of organic solvents (dodecane, benzene and hexanol) and water using this procedure, at several temperatures [30]. The results for the interfacial tension as a function of reduced temperature are shown in Fig. 1, along with the best fit to the scaling function given by eq. (3). First, it is reassuring to find that the predictions of the DPD simulations for the interfacial tension collapse on a single curve despite the different chemical composition of the systems, i.e., there is scaling. Secondly, the scaling exponent obtained from the simulations is $\mu = 1.2$, which is close to the universality accepted value for $3d$ liquids, $\mu = 1.26$.

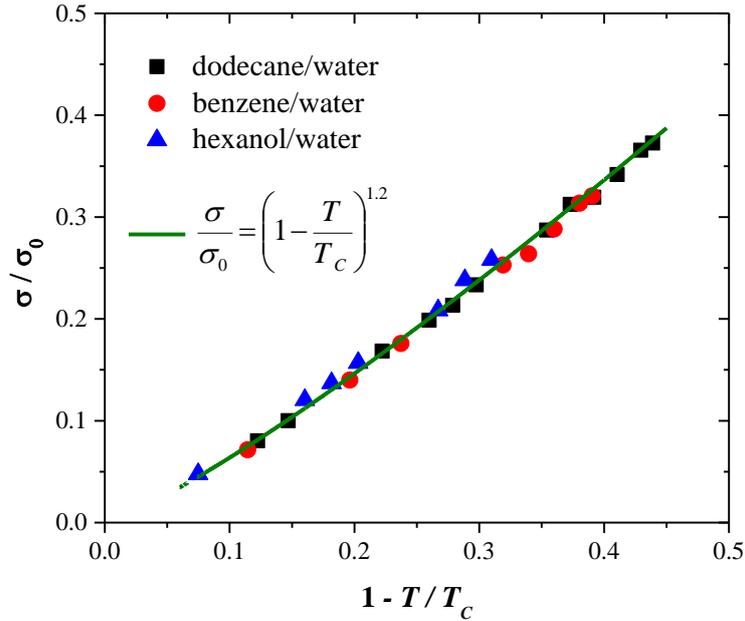

**Fig. 1** Normalized interfacial tension as a function of reduced temperature for three mixtures of organic solvents with water. The solid line represents the best fit to eq. (3), with $\mu = 1.2$. The value of $\sigma_0$ is 81.8 for dodecane/water, 71.9 for benzene/water and 57.3 for hexanol/water. Adapted from ref. [30].

The natural correlation length in these mixtures can be defined as the thickness of the interface between the immiscible liquids, and by tracing its change with varying temperature one can compare with the predictions of eq. (4) [30]. Figure 2 displays the evolution of the correlation length so defined with temperature, for the particular case of the mixture of hexanol (red data) and water (blue data). The correlation length is found to grow with increasing temperature, as expected; in fact it should be infinite when the system reaches the critical temperature, as usual for several critical properties.

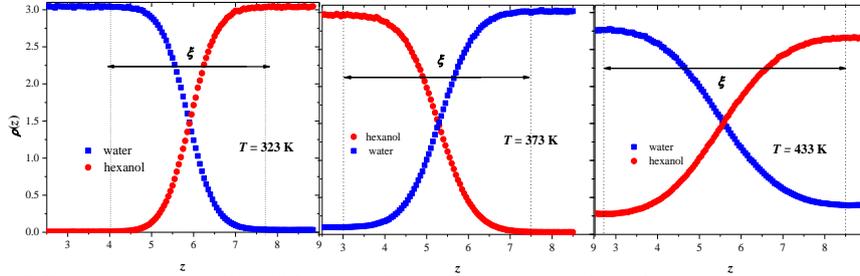

**Fig. 2** Concentration profiles of the water/hexanol interface at three different temperatures. The thickness of the interface, labelled as $\xi$ in the panels shown in the figure, is defined as the correlation length of the system, see eq. (4). Adapted from ref. [30].

Following the same procedure for the mixtures water – benzene and water – dodecane leads to reasonable collapse of all data on a single curve, as seen in Fig. 3. The solid line in Fig. 3 is the fit to eq. (4), with $\nu = 0.63$, in agreement with the value expected for the $3d$ Ising model [28]. The average value for $\nu$ found from the data shown in Fig. 3 is $\nu = 0.67$, hence $\mu = 2\xi = 1.34$. The prediction from eq. (5) for $d = 3$ is $\mu = 2\xi = 1.26$ for the Ising model in $3d$, therefore our simulations confirm Widom's hyperscaling relation, eq. (5), at least for $d = 3$.

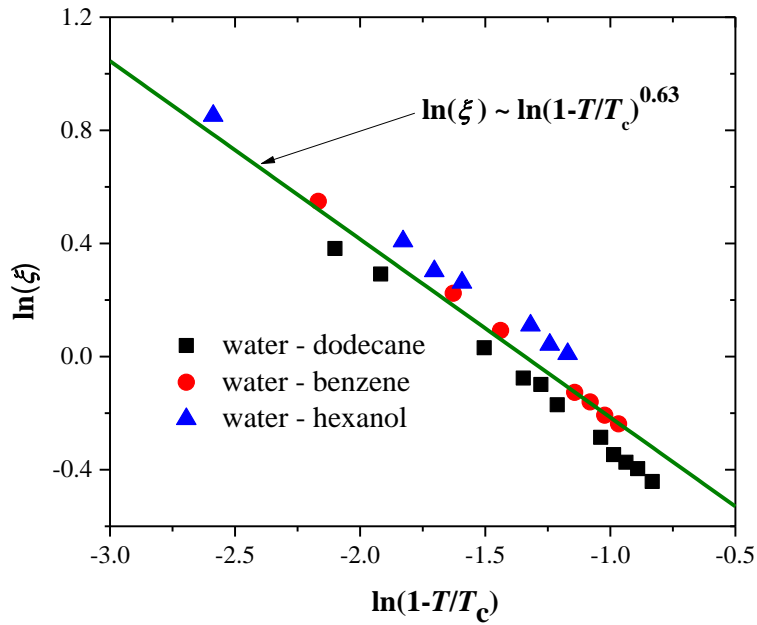

**Fig. 3** Correlation length in mixtures of organic solvents with water, as a function of reduced

temperature. The solid line is the best fit to eq. (4), with critical exponent $v = 0.63$; see text for details. Adapted from ref. [30].

It is important to ask oneself why the interfacial tension of $3d$ liquids modeled with DPD appears to belong to the $3d$ Ising universality class [28]. A simple argument can be provided to supply such an interpretation. The Ising model can be equally applied to spin up/spin down sites (its original purpose) as to occupied/empty sites, or equally well to (site – occupied – by – liquid – 1)/(site – occupied – by – liquid – 2) systems, see illustration in Fig. 4. This is precisely what occurs at the interface between immiscible liquids, as shown by the region where $\xi$ is defined in Fig. 2. In addition to DPD, this equivalence is expected to hold for other interaction models, as long as they are short – ranged, so that next – nearest neighbor interactions can be neglected.

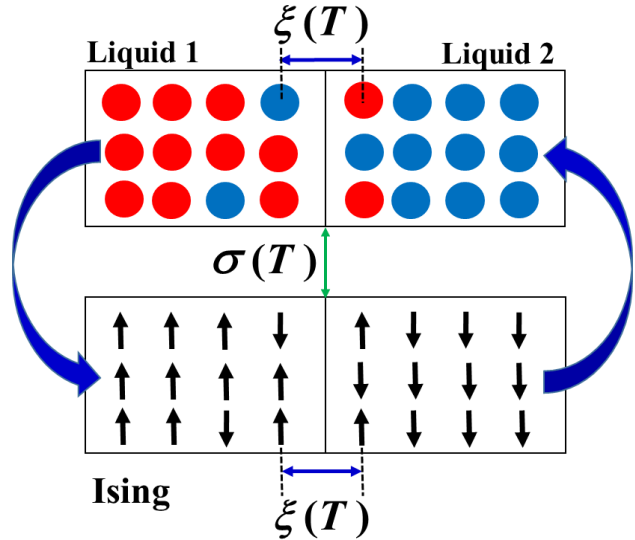

**Fig. 4** An illustrative interpretation of the reason why the interfacial tension between liquids predicted by DPD belongs to the $3d$ Ising universality class. Adapted from the Table of Contents graphic of ref. [30].

## 4    Scaling of the radius of gyration of polyelectrolytes

Early work by Flory and others [1] led to the conclusion that the characteristic length of a polymer in solution, its radius of gyration $R_g$, obeys a scaling law in terms of its polymerization degree, $N$. Such scaling law is given by:

$$R_g = N^v, \tag{6}$$

where $\nu$ is the scaling exponent. Flory arrived at the conclusion that $\nu = 3/(d + 2)$ using insightful yet simple arguments [1]; this relation is found to hold reasonably well in experiments and simulations, except for $d = 3$, where renormalization group calculations yield the universally accepted value $\nu = 0.588$ [31]. Following de Gennes's analogy between polymer statistics and critical phenomena [1], an equivalence can be expressed between the proximity of temperature to the critical point (see eq. (4)), in critical phenomena, and the polymerization degree of large polymers, see eq. (6):

$$N \sim 1/|1 - T/T_C|. \qquad (7)$$

If such analogy holds for the relation between the interfacial tension scaling exponent $\mu$, see eq. (3), and the scaling exponent of a polymer's radius of gyration $\nu$, see eq. (6), then eq. (5) in two dimensions ($2d$) reads simply $\mu = \nu$. Now, if the correlation length of a mixture of immiscible liquids in $2d$ follows Ising's universality class in $2d$, where $\nu = 1$ [28], then $\mu = 1$. If, on the other hand, $\nu$ obeys the scaling expected for the gyration radius of polymers in solution under good – solvent conditions, then $\nu = 3/4$, and applying Widom's hyperscaling relation, eq. (5), one should find that $\mu = 3/4$ also. Research is under way to test these scaling relations and find out which limit applies.

The two leading arguments usually provided to understand scaling properties in polymers are, on the one hand, the self – similar structure of polymers on different scales, and on the other, the absence of long range interactions. The latter is not fulfilled in polyelectrolytes, which are electrically charged polymers. However, experiments [32] and theories [33] have determined that, under special circumstances polyelectrolytes do show scaling characteristics. To help understand recent experiments such as those performed on DNA molecules under changing ionic concentration [34], our group has performed extensive DPD simulations of polyelectrolytes, searching for scaling properties.

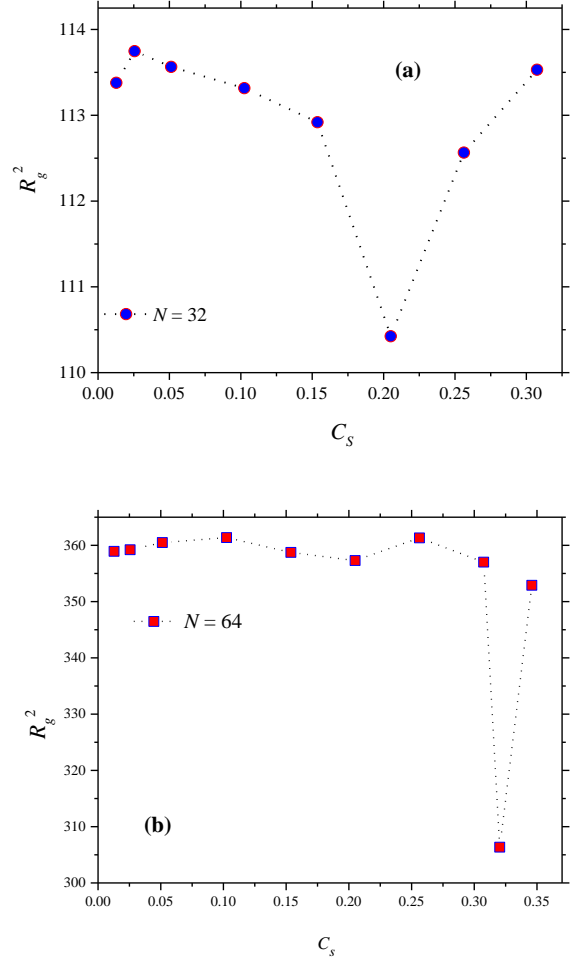

**Fig. 5** Radius of gyration for a polyelectrolyte immersed in a theta solvent as a function of the salt (NaCl) concentration of valence (4:1) for (a) polymerization degree $N = 32$, and (b) for $N = 64$. Axes are shown in reduced DPD units; lines are only guides for the eye. Adapted from ref. [35].

Electrostatics in DPD was introduced first [17] following a simple idea: let every charged particle carry a spatially decaying distribution of charge so that when such distribution is integrated over volume one obtains the full charge carried by the DPD particle. This charge – distribution method was employed to predict the changes in the radius of gyration of a single polyelectrolyte immersed in a theta solvent (all non – electrostatic interactions are equal) while increasing the ionic strength [35]; the results are presented in Fig. 5. For both polyelectrolytes, the short (a) and larger one (b) there appears a minimum in the radius of gyration as the ionic strength is increased, with the minimum being dependent upon $N$. It is however noteworthy that re – expansion of the polyelectrolyte is found when the ionic strength is increased beyond that where the minimum $R_g$ is obtained, regardless the polymerization degree. This phenomenon, which has been observed in experiments [36] and is confirmed by simulations that use interactions different from those used by DPD [37] has been interpreted as being due to charge inversion clouds around the charged monomers on the polyelectrolyte chain [35].

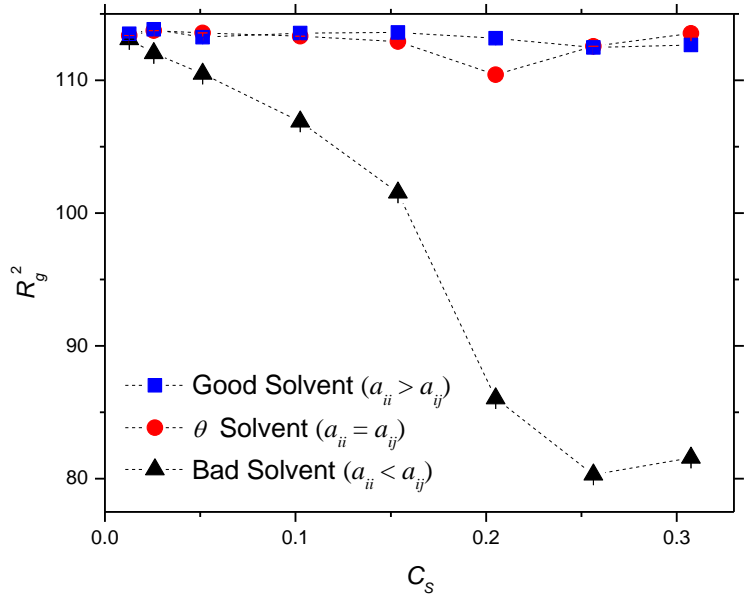

**Fig. 6** Radius of gyration for the polyelectrolyte with $N = 32$ as a function of ionic strength, under conditions of different solvent quality. The latter is modified through the conservative DPD interaction strength, see eq. (1), as indicated in the legend. All quantities are reported in reduced DPD units. Adapted from [35].

Solvent quality is found to play a major role in determining the radius of gyration of polyelectrolytes. Although most simulations and analytic theories are developed for good – solvent conditions under the assumption that is the experimentally relevant case, this assumption is increasingly challenged in several industrial application [38]. Therefore, simulations of the relatively short polyelectrolyte ($N = 32$) were performed for the three solvent conditions to determine its gyration radius at increasing ionic strength; the results can be found in Fig. 6. To change solvent quality one needs only modify the strength of the maximum conservative DPD force, see eq. (1) and the legend in Fig. 6. As expected, the radius of gyration of the polyelectrolyte dissolved in a poor solvent is considerable smaller than in any other case, see black triangles in Fig. 6. The influence of electrostatics is not enough to overcome the short range interactions and full re expansion of the polyelectrolyte after it has collapse is not observed. By contrast, theta – and good – solvent conditions (circles and squares in Fig. 6, respectively) lead to essentially the same radius of gyration. One important difference is that the contraction – re expansion phenomenon found in theta – solvent is less pronounced when the polyelectrolyte is under good – solvent conditions, since by its very definition, it is the solvent that leads to the largest possible polyelectrolyte configuration. The reader is referred to [35] for further details and discussion.

Performing a series of simulations for increasing polymerization degree, it is possible to extract the exponent ν defined in eq. (6), *if* such scaling exists for polyelectrolytes. What is found is that in fact such scaling *does* exist and the value of the scaling exponent can be larger than it is for neutral polymers [1]. Before attempting to understand why scaling behavior is obtained when long range interactions are present, let us first consider a somewhat alternative approach to incorporate electrostatics to the DPD model. As announced in Section 2, here I briefly review very recent work [39] on the use of point charges (rather than charge distributions) in conjunction with Ewald sums, and its application to the prediction of the scaling exponent ν for polyelectrolytes. Let us first recall, see eq. (1), that DPD particles are "soft", thus they can overlap completely. Since the electrostatic interaction blows up when the particles overlap, Groot [17] envisaged a way to prevent this from happening. He argued that using distributions of charge, with an appropriately modified force interaction when particle overlapping began to occur, was sufficient to avoid such artifact. Using distributions of charge means that one must provide a means to solve Poisson's equation for those distributions, which cannot be accomplished exactly, and some ansatz must be used to calculate the full electrostatic interactions and forces. Although this research program is useful and leads to correct results, it suffers from its need to resort to interpolating formulas to solve Poisson's equations. Our approach [39] begins with the realization that DPD particles are only really soft for the smallest coarse – graining degree, i.e., the grouping of one solvent molecule per DPD particle, which is not as useful as larger coarse – graining degrees to model soft matter at the mesoscopic level. Since increasing the coarse – graining degree is tantamount to "hardening" the DPD particles, which in turn makes full particle overlap improbable, it is then unnecessary to use charge distributions and point charges can instead be used. For these (point charges) one can use the full machinery of the Ewald sums, without relying on electrostatic potential energy ansatz [5].

To see how good the point – charge approach can be in DPD, Fig. 7 shows a diagram displaying the percentage of ionic pairs formed when the coarse – graining degree ($N_m$) and point charge strength are increased. The "hardening" of the DPD particle increases if $N_m$ is increased, while increasing the charge requires also of harder DPD particles to avoid the formation of artificial ionic pairs.

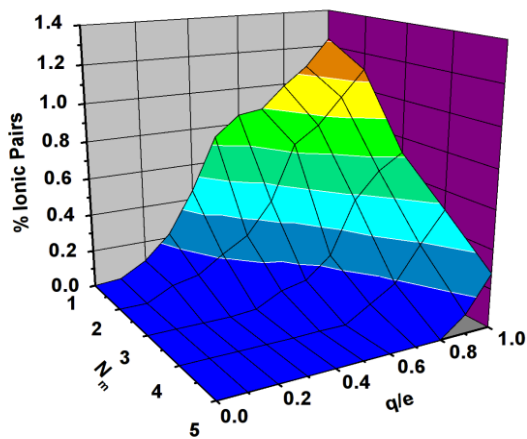

**Fig. 7** Three - dimensional diagram for point charges in DPD, illustrating the formation of ionic pairs, as a function of the coarse – graining degree $N_m$ and the value of the point charge $q$. All simulations were run on GPU's. Adapted from [39].

Using point charges and following a procedure similar to that described in the discussion of Fig. 6, long simulations were carried out for polyelectrolytes of increasing polymerization degree under theta – and good – solvent conditions. The results are to be found in Fig. 8, where only the value of the scaling exponent $\nu$ is reported, for brevity. However, the reader is made aware that each point in Fig. 8 represents a series of simulations for a polyelectrolyte chain of increasing polymerization degree, so that a $R_g$ vs $N$ curve could be generated and the scaling exponent extracted from a linear fit (in a log – log plot). Therefore the need to perform large simulations in relatively short times led us to implement our code so that it could be executed in fast graphical processing units (GPU). The scaling exponent is reported in Fig. 8 as a function of the salt content. The dashed line is the expected value of the scaling exponent for three – dimensional neutral polymers in good solvent, namely $\nu = 0.588$ [31], while the dot – dashed line is Flory´s prediction, $\nu = 0.6$ [1]. For theta – solvent conditions in $3d$ one expects $\nu = 0.5$ for neutral polymers, which is not obtained for polyelectrolytes, as Fig. 8 clearly shows. The scaling for polyelectrolytes in theta solvent is close to the scaling expected for neutral polymers under good – solvent conditions, while polyelectrolytes in good solvent scale like ideal neutral polymers. This can be interpreted as thinking that electrostatic interactions modify solvent quality as well as neutral ones do. That the scaling exponent is larger for polyelectrolytes than for polymers may perhaps be expected also, since electrostatic repulsion between neighboring charged monomers along the polyelectrolyte chain would lead to a more stretched out configuration. What might not be obvious is why scaling is obtained despite the presence of electrostatics, as Fig. 8 shows. Once polyelectrolytes are introduced, even at zero ionic strength, counter ions must also be added to the system, to keep it globally neutral. Those counter ions tend to group around charged monomers in the polyelectrolyte, screening the charges in it and effectively reducing the range of the electrostatic interaction. Then, the same arguments used for scaling in neutral polymers can be recalled. This is admittedly an overly simplified view, one that does not take into account important aspects such as the change in the persistence length with charge, Kuhn's length, Manning's condensation, and the solvent's permittivity change with charges [40]. Although important for a full understanding of polyelectrolytes in solution, those aspects are not as fundamental in yielding scaling characteristics as the shortening of the interaction length, while preserving self – similarity on different scales. The rest of this chapter is devoted to results for neutral systems.

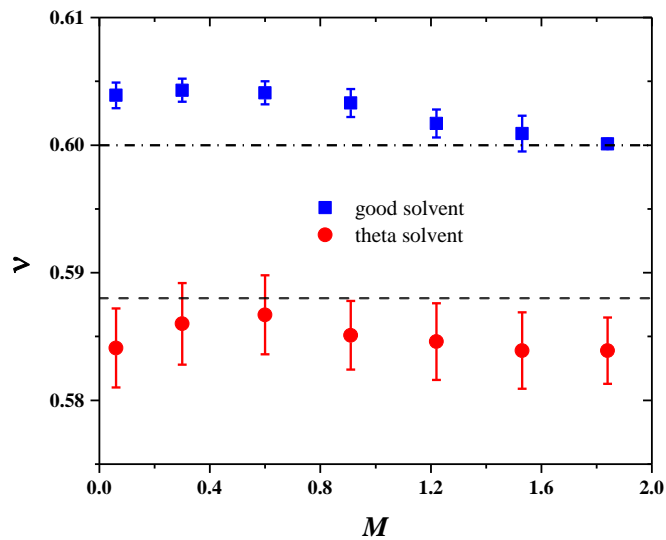

**Fig. 8** The scaling exponent of the radius of gyration as a function of polymerization degree, $\nu$, see eq. (6) obtained for a polyelectrolyte chain of increasing size, as a function of increasing ionic strength ($M$). The dashed and dot – dashed lines are included as reference to the currently accepted value for neutral polymers in good solvent in 3d (n = 0.588) and Flory's prediction. All simulations were performed on GPU cards, using the SIMES code [42]. Adapted from [39].

## 5  Scaling properties of polymer brushes in equilibrium

When polymer chains are grafted on a surface and the grafting sites are closer to each other than the radius of gyration of the polymers in solution, they form arrangements that resemble brushes. They are known to display some scaling properties that depend on parameters such as grafting density, polymerization degree, and of course solvent quality. They are also important from a technological point of view, since polymer brushes can be very effective as colloidal dispersion stabilizers when used as coats on colloidal particles, such as in paints [12]. Polymer brushes also constitute excellent lubricating tools, able to reduce friction between brush – coated surfaces by three – orders of magnitude [42]. Since those works are well known [43] I shall focus here on a less known but increasingly important type of polymer brushes: biological polymer brushes. One particular example that draws our attention is the case of biological brushes on the surface of healthy and cancerous human cervix epithelial cells. Various types of physical probes have been able to detect linear protuberances in the form of brushes covering the surfaces of cells [44]. Those brushes are complex entities, described as microvilli, microridges and cilia, thought to be made of filaments of acting. Experiments using atomic force microscopy (AFM) detected different mechanical response between the surface of cancerous and healthy cells [45], with those differences being attributed to the brushes.

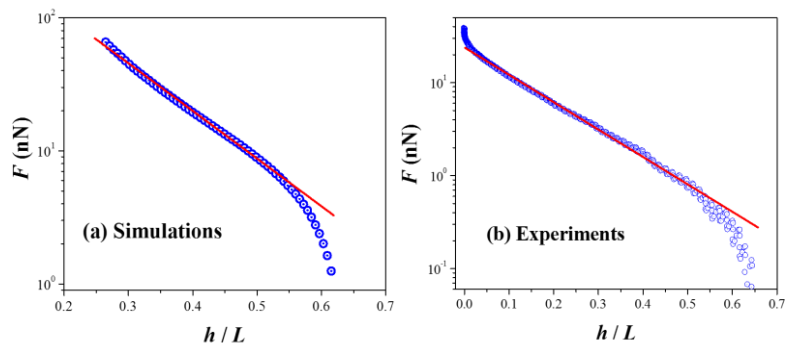

**Fig. 9** Force applied by an AFM probe to the surface of normal cervical epithelial cells covered by biological brushes as predicted by DPD simulations [ref AGG-SJAG] (a), and determined experimentally [45], (b). The solid line is the best fit to the Alexander – de Gennes's brush scaling law, see text for details. The $x$ –axis was normalized by the average size of the unperturbed brushes. Adapted from the supplementary information of ref. [46].

Using DPD simulations, most of those results have been interpreted and are now understood [46]. In particular, the force exerted by the mesoscopic tip of an AFM on the surface of healthy epithelial cervical cells covered with biological brushes has been correctly predicted, as shown in Fig. 9. Notice also that both experiments and numerical predictions are found to be in agreement with the polymer brush scaling law proposed by Alexander and de Gennes, which can be written as

$$F \sim F_0 e^{-2\pi h/L}, \qquad (8)$$

when $0.2 \leq h/L \leq 0.9$, see [47]. In eq. (8) $F_0$ is a constant expressed in terms of the thermal energy, the radius of curvature of the AFM tip, and the brushes grafting density [45]. The force decays linearly on a semi log scale, with slope proportional to the reduced brush thickness ($h/L$), as Fig. 9 illustrates. The physical arguments that lead to the scaling predicted by eq. (9) can be stated as follows: as the brush is compressed, the local osmotic pressure is increased. This contribution must compete with the attractive elastic energy stored in the polymer chains, leading to the scaling law in eq. (8) [1].

This accuracy of DPD simulations in capturing scaling behavior in polymer brushes is not accidental as it has been found to be successful in other applications, see for example [22]. Additional examples are displayed in Fig. 10.

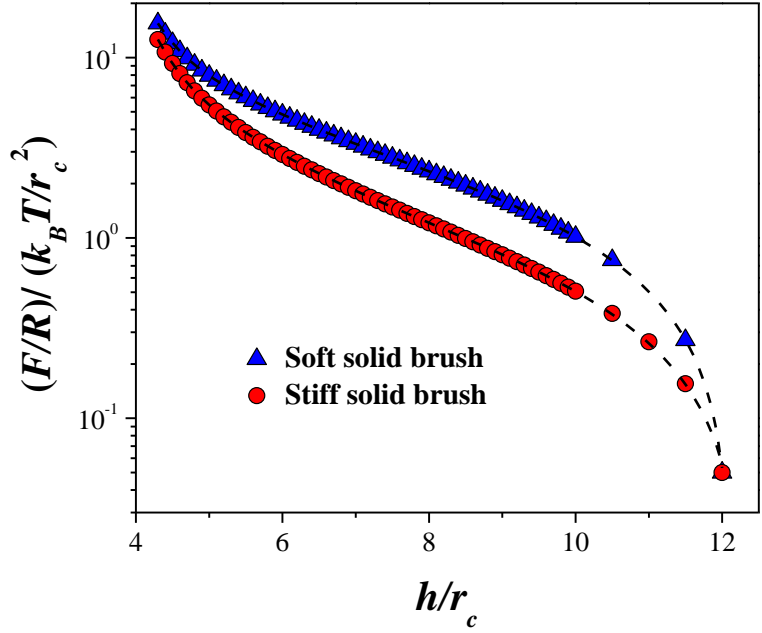

**Fig. 10** Force applied by a mesoscopic sized tip of an AFM per its radius of curvature, as a function of the distance between the surface of cancerous epithelial cervical cells and the AFM's tip, obtained from DPD simulations [46]. Solid triangles correspond to brushes made up of chains where each monomer is joined to its neighbors by soft springs. Solid circles correspond to brushes with stiff springs; chains' heads in both soft and stiff brushes are fixed on the surface of the cell, thereby their label as "solid". Dashed lines are the fits to the Alexander – de Gennes scaling adapted to a three – length brush, as found in experiments [45], see text for details. Adapted from the supplementary information of ref. [46].

The results of DPD simulations presented in Fig. 10 (solid symbols) along with their fit to the Alexander – de Gennes scaling law (dashed lines) correspond to a model cancerous epithelial cell covered by brushes of three different lengths and grafting densities [46]. The model was constructed following experimental findings on that type of cell, which showed that cancerous cervical epithelial cells were covered by a thick small brush, coexisting with less dense, medium sized chains; the micrographs also displayed a rare, much less dense and much larger third brush [45]. Fitting the surface force to a three – brush Alexander – de Gennes scaling force, given by

$$\frac{F}{R} / \left(\frac{k_B T}{r_c^2}\right) = A_1 e^{-(x-x_0)/L_1} + A_2 e^{-(x-x_0)/L_2} + A_3 e^{-(x-x_0)/L_3} + B \quad (9)$$

yields the dashed lines shown in Fig. 10. Obviously, the predictions are in excellent agreement with the scaling and with the experimental trends. In eq. (9) the lengths of the three different brushes are labelled $L_i$, $x_0$ is the complete three – length brush maximum compression, and $A_i$ and $B$ are adjustable constants [46]. The values of all these parameters obtained from regression analysis are also in excellent agreement with the model parameters used in the simulations; for full details the reader is referred to [46].

It is satisfying to find that physical models such as the DPD model and its thermostat are successful in reproducing complex phenomena occurring in many – body systems, such as those representing biological brushes. The fundamental reasons behind this success are found in the DPD model interactions, which despite the fact they are short range, they still lead to a non – vanishing second virial coefficient [24]. This sets the DPD model ahead of mean – field theories, where chain – chain interactions and even solvent interactions are neglected. There are weak but finite DPD interactions between chains in brushes, an aspect that several popular scaling theories neglect [48], one that is fundamental in capturing many – body collective phenomena, as typically required for scaling laws to work.

## 6  Scaling properties of polymer brushes under flow

Scaling laws of dynamical properties of polymer brushes have been developed, some of which are briefly discussed in this section, particularly with focus on brushes under stationary, Couette flow [49]. Important technological applications demand fundamental knowledge of the properties of polymer brushes under flow, such as those where brushes are used as friction reducing agents [50]. Basic research in this direction has helped the plastic industry design better and more efficient schemes for the production of plastic bags, for example, so that users can separate the sheets of their plastic bags more easily when they go grocery shopping [51]. No less important is the basic understanding of the mechanisms that give rise to scaling trends even when polymer brushes are subjected to flow. In what follows I show some results our group has obtained regarding the scaling of dynamical properties of polymer brushes, with focus on two measurable quantities: viscosity of a fluid composed of polymer brushes under flow, and the friction coefficient between those opposite brushes and the solvent.

The setup of the simulations reported in this section is as follows. Polymer chains are grafted by one of their ends to parallel surfaces placed at the ends of the simulation box in the $z$ – direction, see Fig. 11. Those ends, represented by blue beads in Fig. 11, are subjected to an external force that makes them move to the right (top surface in Fig. 11) with constant velocity ($v_0$), while the beads grafted to the bottom surface move to the left with constant velocity $-v_0$. This setup leads to a linear velocity gradient for all particles confined within the pore of width $D$ shown in Fig.11; this type of stationary flow is known as Couette flow [49]. The shear rate in this situation is defined as $\dot{\gamma} = 2v_0/D$, which is constant since the velocity of the grafted beads and the spacing between the surfaces are constant. The viscosity can then be expressed as follows:

$$\sigma = \eta \dot{\gamma}, \tag{10}$$

where $\sigma = \langle F_x \rangle / A$ is the shear stress on the sample, given by the mean force on the particles along the $x$ – direction divided by the transversal area of the surfaces, see Fig. 11.

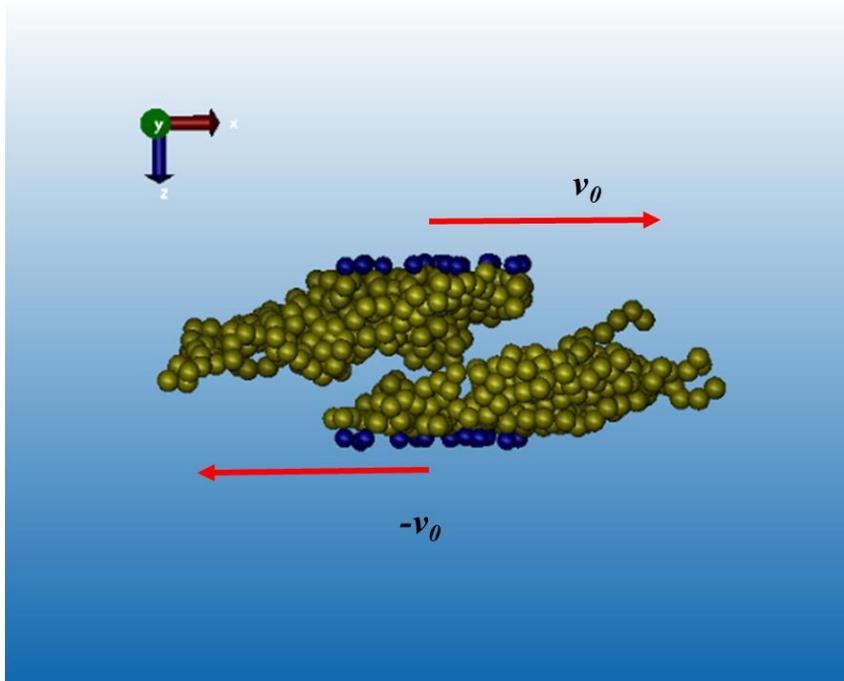

**Fig. 11** Snapshot of the simulation cell used to model polymer brushes under stationary flow. Blue beads represent the ends of the brushes grafted on the surfaces (not shown), which are moved by an external force so that they move at constant speed $v_0$ to the right (upper surface) or to the left (lower surface). The rest of the polymer chains are shown in ochre beads; the solvent particles are not shown for simplicity.

The coefficient of friction can also be obtained from the simulation setup shown in Fig. 11, as the ratio of the average force along the $x$ – direction over the average force along the $z$ – direction, i.e. perpendicularly to the wall, as given by eq. (11):

$$\mu = \langle F_x \rangle / \langle F_z \rangle. \tag{11}$$

Non Newtonian fluids, and polymer fluids usually belong to this category, are those with viscosities that depend on the applied shear rate, as given by eq. (10). Shear thinning behavior is found in many polymer liquids, namely their viscosity is reduced when shear rate is increased. There is a critical shear rate, $\dot{\gamma}^*$, below which the fluid behaves as a Newtonian fluid. For values larger than $\dot{\gamma}^*$ shear thinning behavior sets in. In fact, it is possible to define a universal dimensionless number, the so – called Weissenberg number [52], $We$, given by $We = \dot{\gamma}/\dot{\gamma}^*$ so that $We \geq 1$ signals non Newtonian behavior.

Figure 12 shows the results of several DPD simulations, performed on brushes of different polymerization degree, subjected to increasing flow under theta solvent conditions. The idea behind those simulations was to determine if characteristics such as increasing polymer length eventually lead to size – free physical trends. Let us first focus on panel (a) in Fig. 12, where the dynamic viscosity of brushes – at the same grafting density for all cases reported in Fig. 12 – is shown as a function of increasing Weissenberg number, $We$. Predictions of simulations of polymer brushes in theta – solvent conditions under increasing shear flow rate show that scaling behavior *is* obtained for the viscosity of polymer brushes regardless the polymerization degree, where the scaling exponent is found to be equal to $\zeta = -0.31$ on a log – log scale, see Fig, 12(a).

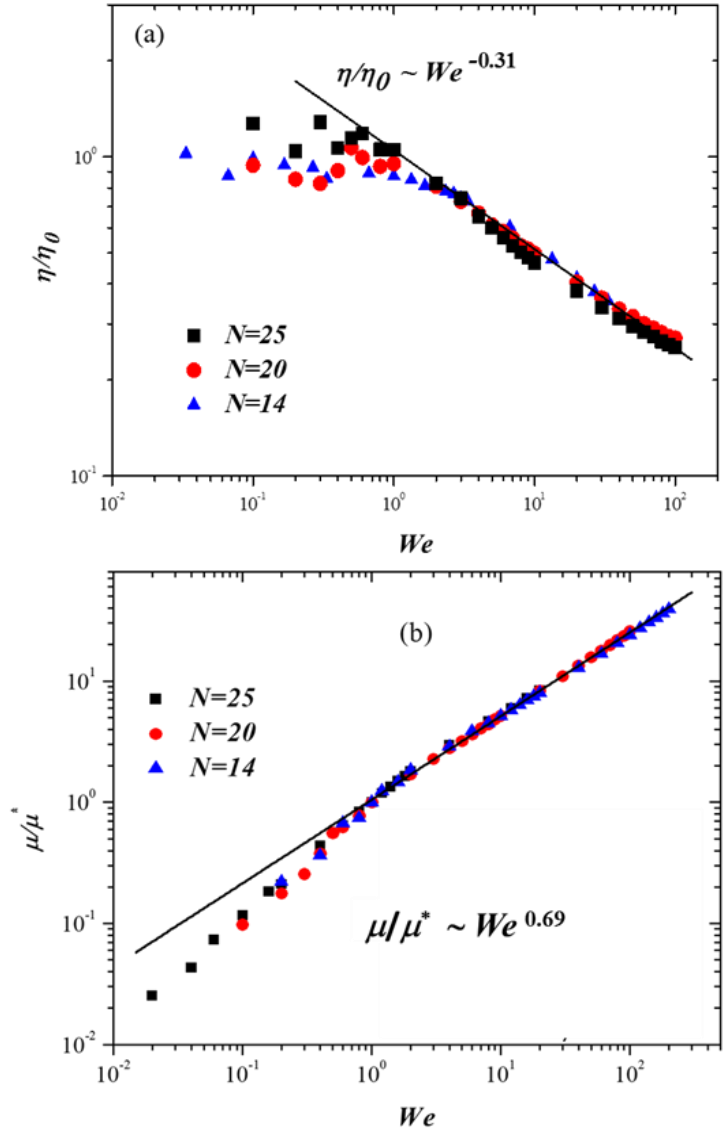

**Fig. 12** DPD non – equilibrium simulations of polymer brushes. (a) Reduced dynamical viscosity as a function of the Weissemberg number (*We*) for polymer brushes of three different lengths, as indicated by the legend. The solid line represents the fit to the scaling model proposed by Galuschko et al. [52]. (b) Reduced friction coefficient between brushes of increasing polymerization degree as a function of *We*. Notice scaling behavior is captured for both properties when *We* > 1, as indicated by the solid lines. Adapted from [54].

The reduced coefficient of friction for the same polymer brushes as those reported in Fig. 12(a) is shown in Fig. 12(b), as a function of $We$. Once again, scaling trends independent of the polymerization degree are obtained once $We \geq 1$, with the scaling exponent being $\kappa = 0.69$. Remarkably, universal curves are obtained for all polymerization degrees modeled [52], and a relationship can be established between the scaling exponents of the viscosity and the friction coefficient, yielding the equation [52]:

$$\kappa - \zeta = 1, \qquad (12)$$

where $\kappa = 0.69$ and $\zeta = -0.31$ for polymers under theta – solvent conditions [54]. Simulations carried out by other groups using different model interactions for polymers under good solvent conditions yield values for these scaling exponents given by $\kappa = 0.57$, and $\zeta = -0.43$ [52], which are clearly different from those obtained for brushes in theta solvent. However, it is most remarkable that eq. (12) is equally fulfilled, regardless the solvent quality. Scaling behavior is obtained for polymer brushes under strong confinement ($D \ll R_g$) and Couette flow ($We \gg 1$) because chains are strongly stretched along the shear direction, giving rise to a situation where flow plays the role of an external field whose role is mostly brush alignment [52]. Therefore, for large $N$ and under strong flow and confinement, polymers appear again to be self – similar and scaling ensues. Under those conditions, the brushes whose scaling is reported in Fig. 12 behave like polymer melts, for which the scaling exponent $\nu$, see eq. (6) is $\nu = 0.5$. Then, it can be shown [52] that $\mu/\mu_0 = \langle F_x(\dot{\gamma}) \rangle / \langle F_x(\dot{\gamma}^*) \rangle = N/N^{-0.5}$, and that $N \sim We^{6/13}$ which combined yield $\mu/\mu_0 = We^{9/13} = We^\kappa$, or $\kappa \approx 0.69$; using eq. (12) one gets $\zeta = -0.31$, in excellent agreement with the results shown in Fig. 12 [54].

Before leaving this section I comment briefly on the scaling of polymer brushes under flow when there are also free chains of the same type of polymers that make up the brushes. This is not only of academic interest but it is also important in plastic sheet production, where polymers are injected into the plastic matrix (which later will constitute the plastic sheet) so that they migrate to the surfaces of the matrix during the extrusion process. As the plastic cools, more and more polymer chains migrate, forming brushes, thereby reducing friction between sheets and energy consumption. However, some chains get desorbed and get trapped between opposite sheets, creating a complex confined fluid, one that includes free polymer chains, polymer brushes, and solvent particles. The question then arises as to whether those free chains help reduce the coefficient of friction (COF) or not. Figure 13 illustrates the process just described, where the COF and the viscosity are reported also. Notice that in this case the variable is not the shear rate but the polymer chains' grafting density, $\Gamma$ [55]. The results shown in the center of Fig. 13 have been found to be in excellent agreement with experiments performed in both academia and in the private sector [56], but since the focus of this chapter is on scaling properties, I invite the interested reader to review reference [55], and references therein.

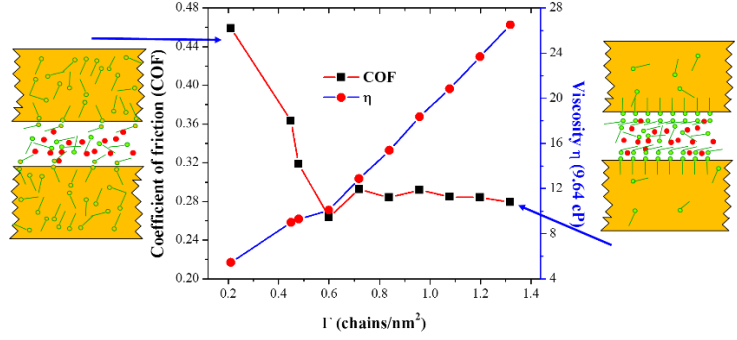

**Fig. 13** Schematics of the polymer chains migration (green molecules) to the surfaces of the plastic matrix (orange colored sheets), where they form brushes as the plastic cools. Some chains are desorbed and form free aggregates, interacting with the brushes and the solvent (red particles). Couette flow is applied and the COF (black squares) and viscosity (red circles) are calculated at increasing brush grafting density ($\Gamma$). Adapted from the Table of Contents Graphic of ref. [55].

In Fig. 14 I show the average force along the direction perpendicular to the surfaces on which polymer chains are grafted, as a function of their grafting density under theta solvent conditions. Notice the scale on both axes is logarithmic. The squares are results of DPD simulations where only brushes are subjected to constant Couette flow; the circles correspond to DPD simulations of brushes and free polymer chains under flow. It is tempting to apply the Alexander – de Gennes (AdG) scaling law to this case, even though it was not derived for brushes under flow. For the present purposes, let us write the AdG law as follows:

$$\langle F_z \rangle = A k_B T f(a, D, N) \Gamma^y \ , \tag{13}$$

$$y = \frac{3\nu}{3\nu - 1}. \tag{14}$$

In eq. (13) $A$ is the surfaces' transversal area, and $f$ is a function that depends on the polymer's monomer size, $a$, the distance between surfaces, $D$, and the polymerization degree, but it does not depend on $\Gamma$; therefore it is not shown here, for simplicity. The exponent $y$ depends on the familiar exponent $\nu$. For theta – solvent polymers, $\nu = 0.5$ and $y = 3$. Another scaling theory, proposed recently, is that of Kreer and Balko's (KB) [53], which predicts that

$$\langle F_z \rangle = A k_B T g(a, D, N) \Gamma^{y'} \ , \tag{15}$$

$$y' = \frac{2 + 5\nu}{3(3\nu - 1)}. \tag{16}$$

The function $g$ in eq. (15) depends on the same variables as function $f$ in eq. (13), although they are different. However, their explicit form is not relevant to the present discussion. What is important is to note that $y$ (eq. (14)) and $y'$ in eq. (16) are both equal to three for $\nu = 0.5$, even though they are based on different assumptions. The dashed line in Fig. 14 is the fit to a function $\sim \Gamma^3$, which indicates that there is scaling for these systems, and that the presence of free chains does not change the scaling properties of the compression force of polymer brushes in a theta solvent.

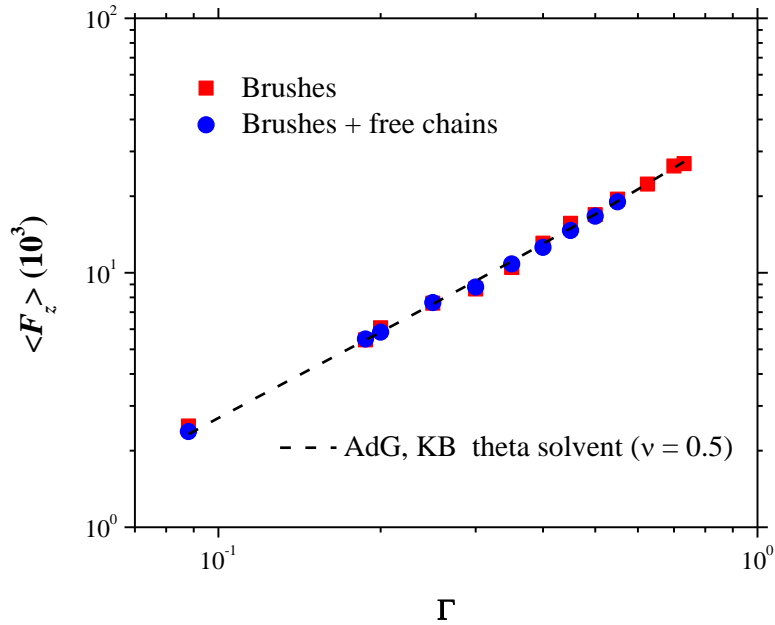

**Fig. 14** Average force along the direction perpendicular to the confinement of polymer brushes (red squares), and brushes plus free chains (blue circles) as a function of polymer grafting density, under constant Couette flow, obtained using DPD simulations. The dashed line is the fit to the Alexander – de Gennes (AdG) [47] and Kreer – Balko (KB) [53] scaling laws, see eqs. (13) and (15), respectively. Adapted from [55].

The fact that eqs. (13) and (15) lead to the same scaling exponent means that chain – chain interaction and brush interdigitation (allowed by KB but not by AdG) are not the principal factors that give rise to scaling, under theta conditions. Other aspects define the scaling for these systems, such as the compression ($D$) and the polymerization degree. The fact that both systems (only brushes and brushes plus free chains) are under flow does not affect the scaling either. The physical reason is to be found in the fact that $D$ does not change. Differences in scaling behavior between equilibrium and non – equilibrium simulations of polymer brushes are expected to occur for the force along the $x$ – direction, but that is beyond the scope of the present work.

# 7 Scaling in lower dimensions

Lastly, I present some results for scaling of polymers in two dimensions (2*d*), with particular emphasis on the scaling of their disjoining or solvation pressure Π [57]. When fluids are confined, the component of the pressure tensor along the direction normal to the confinement ($P_N$) is in general different from other components; in particular, it is different from the unconfined, bulk pressure ($P_B$). Such difference is precisely Π; it is important because it can be used as a means to gauge colloidal stability (Π is proportional to the free energy difference between the bulk and confined systems), for example [58]. It can be measured using a surface force apparatus, or with AFM [47]. In systems with varying concentration of a given component, the disjoining pressure should be proportional to the osmotic pressure ($\pi$). The latter was shown a long time ago [59] to obey the following scaling relation:

$$\pi \sim c^{\frac{\nu d}{\nu d - 1}}, \tag{17}$$

where *c* is the monomer concentration, *d* the spatial dimension, and $\nu$ the scaling exponent of the Flory radius. Given the fact that $\pi$ and Π are related, one should expect that the latter obeys a scaling law as well.

For polymer chains under strong confinement along the z – direction, see schematic representation in the inset in Fig. 15, the chains are effectively restricted to move on a quasi – 2*d* plane. Following arguments similar to those used by des Cloiseaux to obtain eq. (17) [59], but with the most important difference that the separation between surfaces (*h*) remains small but finite, the author and E. Pérez showed [60] that Π scales as

$$\frac{\Pi}{k_B T} \sim c^{2\nu_{2d}/(2\nu_{2d} - 1)} \tag{18}$$

where $\nu_{2d}$ represents the value of the ubiquitous Flory's scaling exponent $\nu$ in 2*d* (see eq. (6)), whose value depends not only on *d* but also on solvent quality. The scaling of Π proposed in ref. [60] and shown in eq. (18), the first ever reported for the disjoining pressure, was tested with DPD simulations of a fixed number of polymer chains under increasing confinement (reducing *h*, see diagram in Fig. 15). By reducing the spacing between the surfaces, the volume is reduced and monomer concentration (*c*) can be increased. This renders the simulation of several polymer concentrations unnecessary, since with a fixed number of chains one can sweep over several monomer concentration values.

When the distance between the surfaces is reduced, the fluid confined may not be in the same state of thermodynamic equilibrium, unless the chemical potential between it and the surrounding fluid is kept constant. This restriction means that simulations of confined fluids must be performed in the so – called Grand Canonical ensemble, where in addition to volume and temperature, the chemical potential must also be kept constant [61]. Failure to do some may lead to vastly different predictions between constant – density simulations and constant chemical potential simulations, see [62]. Figure 15 shows the results of Monte Carlo simulations in the Grand Canonical ensemble for a fluid containing polymer chains and solvent particles, interacting through DPD forces [19], under theta solvent conditions. Only solvent particles are exchanged between the confined fluid and the virtual reservoir.

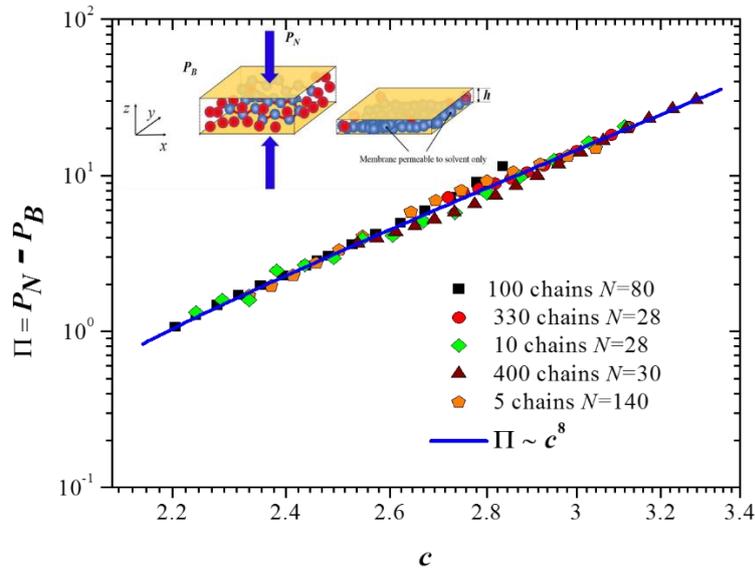

**Fig. 15** Grand Canonical Monte Carlo simulations of polymer confined on quasi 2d – space using the DPD interaction model. The $y$ – axis is the disjoining pressure and the $x$ – axis represents the polymers' monomer concentration. The symbols represent data for chains of various polymerization degrees, as indicated by the legend. The solid line is the fit to eq. (18) with $\nu_{2d} = 4/7$; see text for details. Both axes are reported in reduced DPD units. The cartoon in the upper left corner shows the simulation setup; red beads are solvent particles and blue beads are monomers that make up the chains. Adapted from [60].

The different symbols in Fig. 15 correspond to chains to increasing polymerization degree, ranging from $N = 28$ up to $N = 128$; once again it is quite remarkable to find all data collapse into a single curve, signaling that scaling occurs. The solid line in Fig. 15 corresponds to the best fit to the function $\Pi \sim c^8$; comparing this exponent with the one predicted by eq. (18) yields $\nu_{2d} = 4/7$. This is precisely the same value predicted by other scaling theories for two – dimensional polymers under theta – solvent conditions [63, 64]. Simulations performed for quasi – 2d chains under good solvent conditions (not reported here for brevity) of the disjoining pressure for polymers of various polymerization degrees yield the commonly accepted value, $\nu_{2d} = 3/4$ [1]. Therefore, the scaling law predicted by Gama Goicochea and Pérez [60], eq. (18) is very robust. For more discussion about the implications of this scaling, as well as full for details of its derivation and of the simulations whose results are shown in Fig 15, see ref. [60]. For fractal scaling of cluster aggregation, see for example ref. [65].

## 8   Conclusions

The physicist Thomas A. Witten once wrote "*Why should a physicist be interested in polymers? They do not hold the key to vast resources of energy like atomic nuclei. They do not defy intuition with ultrasmall dissipation like superconductors and superfluids. They do not reveal subtle new nonabelian symmetries as do subatomic particles. Nor do they hold secrets about the origin or fate of the universe*" [66]. This is indeed true, yet as Witten himself goes on to argue in [66], polymers constitute a most important field of study, for polymer liquids display properties that can be understood using powerful analogies with critical phenomena; they have motivated the development of sophisticated experimental techniques; they are commonplace in modern society, and if that was not enough, even biological matter can be thought of and understood as polymer liquids.

The study of polymers fluids, and soft condensed matter in general, has benefited in recent decades from numerical simulations, which are ever faster, adaptable to model systems of ever increasing complexity and, with costs of powerful processors becoming more competitive, accessible to a wider number of scientists worldwide. Here I have focused on reviewing the modeling of scaling properties of soft matter systems using in particular the technique known as dissipative particle dynamics, carried out by our group. However, all scaling properties reported here are independent of the technique used; some have been obtained by other groups using different models or measured in various experiments. The fact that the same scaling exponents are found in chemically different compounds, using vastly different techniques is gratifying for those using such techniques, and it is also a beautiful example of the unifying concept of scaling in physics.

### Acknowledgments

The author wishes to thank his collaborators, with whom most of the results reported here were obtained, in particular: F. Alarcón, S. J. Alas Guardado, M. A. Balderas Altamirano, J. Barroso – Flores, R. Catarino Centeno, J. S. Hernández Fragoso, J. D. Hernández Velázquez, J. Klapp, R. López – Esparza, R. López – Rendón, E. Mayoral, S. Mejía – Rosales, C. Pastorino, R. Patiño Herrera, E. Pérez, G. Pérez – Hernández, Z. Quiñones, E. Rivera – Paz, K. A. Terrón – Mejía, J. Vallejo and M. A. Waldo. Educational discussions with E. Blokhuis and I. Sokolov are also gratefully acknowledged. For computational resources the author is indebted to ABACUS, where some calculations were run; to the high performance cluster Yoltla at UAM – Iztapalapa; to Universidad de Sonora for access to the Ocotillo cluster at their High Performance Computational Area; to the CNS supercomputing facilities at IPICyT, to the Olinka cluster at UAEM, and to the Laboratorio Nacional de Caracterización de Propiedades

Fisicoquímicas y Estructura Molecular Supercómputo Universidad de Guanajuato. For technical support at the IFUASLP, J. Limón is also acknowledged. This work was supported in part by project Proinnova – CONACYT, through grant 231810.